\documentclass[conference]{IEEEtran}

\usepackage{amsmath,mathtools}
\usepackage{amssymb}
\usepackage{pifont}
\usepackage{tikz}
\usetikzlibrary{positioning, arrows.meta, fit, shapes}

\usepackage{graphicx}
\usepackage{amsmath}
\usepackage{algorithm}
\usepackage{subfig}

\usepackage{algpseudocode}
\usepackage{boxedminipage}
\usepackage{fancyhdr}
\usepackage{multirow}

\usepackage{array}
\usepackage{textcomp}
\usepackage{stfloats}
\usepackage{csquotes} 
\usepackage{url}
\usepackage{verbatim}
\usepackage{cite}
\usepackage{eqnarray}
\def\BibTeX{{\rm B\kern-.05em{\sc i\kern-.025em b}\kern-.08em
    T\kern-.1667em\lower.7ex\hbox{E}\kern-.125emX}}
\usepackage{balance}
\usepackage{url}

\usepackage[letterpaper, top=1in, bottom=1in, left=1.015in, right=1.015in]{geometry}
\begin{document}

\title{A Modular KDN-Based Framework for IT/OT Autonomy in Industrial Systems}

\author{
    \IEEEauthorblockN{Tuğçe BILEN\IEEEauthorrefmark{1} and Mehmet OZDEM\IEEEauthorrefmark{2}}
    
    \IEEEauthorblockA{\IEEEauthorrefmark{1}Department of Artificial Intelligence and Data Engineering\\Faculty of Computer and Informatics\\
    Istanbul Technical University, Istanbul, Turkey\\}
    \IEEEauthorblockA{\IEEEauthorrefmark{2}Turk Telekom, Istanbul,Turkey \\
    Email: bilent@itu.edu.tr, mehmet.ozdem@turktelekom.com.tr}
}

\maketitle

\begin{abstract}
The convergence of Information Technology (IT) and Operational Technology (OT) is a critical enabler for achieving autonomous and intelligent industrial systems. However, the increasing complexity, heterogeneity, and real-time demands of industrial environments render traditional rule-based or static management approaches insufficient. In this paper, we present a modular framework based on the Knowledge-Defined Networking (KDN) paradigm to enable adaptive and autonomous control across IT/OT infrastructures. The proposed architecture is composed of four core modules: {Telemetry Collector}, {Knowledge Builder}, {Decision Engine}, and {Control Enforcer}. These modules operate in a closed control loop to continuously observe system behavior, extract contextual knowledge, evaluate control actions, and apply policy decisions across programmable industrial endpoints. A graph-based abstraction is used to represent system state, and a utility-optimization mechanism guides control decisions under dynamic conditions. The framework’s performance is evaluated using three key metrics: {decision latency}, {control effectiveness}, and {system stability}, demonstrating its capability to enhance resilience, responsiveness, and operational efficiency in smart industrial networks.
\end{abstract}

\begin{IEEEkeywords}
IT/OT Integration, Knowledge-Defined Networking (KDN), Autonomous Network Control, Industrial Networks, Smart Manufacturing.
\end{IEEEkeywords}

\thispagestyle{fancy}

\pagestyle{fancy}
\fancyhf{}
\fancyhead[C]{\scriptsize Accepted by Workshop on Systems and Technologies for IT/OT Integration in Industry 4/5. 0 Environments in IEEE 30th International Workshop on Computer Aided Modeling and Design of Communication Links and Networks (CAMAD), ©2025 IEEE}
\renewcommand{\headrulewidth}{0pt}

\section{Introduction}
The convergence of Information Technology (IT) and Operational Technology (OT) is becoming increasingly critical in the development of autonomous and intelligent industrial systems \cite{10811278}, \cite{BOZKAYA2023103254}. While the IT layer enables centralized computation, data analytics, and wide-area communication, the OT layer governs real-time control, machine coordination, and process-critical operations within production environments. Integrating these two domains allows industrial networks to achieve greater flexibility, scalability, and automation, enabling adaptive responses to changing conditions and unlocking new levels of operational intelligence \cite{10478382}, \cite{9927253}.

Despite these advantages, seamless IT/OT integration introduces significant challenges. As industrial networks grow in scale and complexity, traditional rule-based or static management approaches are no longer sufficient to ensure the reliability, efficiency, and adaptability required in such dynamic environments. These methods lack the ability to interpret contextual data, adapt to unforeseen conditions, or autonomously manage heterogeneous devices and workloads. Bridging the gap between IT and OT layers demands a new class of management frameworks capable of real-time observation, contextual reasoning, and autonomous control across distributed and mission-critical infrastructures.

In this paper, we propose a modular, AI-enhanced framework based on the Knowledge-Defined Networking (KDN) paradigm to enable autonomous IT/OT integration. The proposed architecture introduces a layered intelligence mechanism into the industrial network stack. It includes a \textit{Telemetry Collector}, which continuously gathers real-time metrics from both IT and OT endpoints, including traffic patterns, link status, and system-level events. These raw data streams are processed by the \textit{Knowledge Builder}, which transforms them into structured, context-aware knowledge representations using statistical modeling and graph-based encoding. This knowledge feeds into the \textit{Decision Engine}, which leverages AI and optimization techniques to select optimal control actions under dynamic constraints. Finally, decisions are enforced through the \textit{Control Enforcer}, which translates high-level instructions into low-level commands executed via SDN interfaces and industrial protocol adapters. The main contributions of this work are summarized as follows:
\begin{itemize}
  \item We propose a modular KDN-based architecture tailored for real-time IT/OT convergence in industrial environments.
  \item We design a knowledge-driven decision engine that applies AI and optimization techniques to support autonomous control under uncertainty.
  \item We define a layered control loop for closed-loop observability and enforcement, improving adaptability and scalability across heterogeneous systems.
\end{itemize}

\subsection{Related Work}
Several studies have explored frameworks and methodologies to address the challenges of IT/OT integration in industrial networks. Below, we review key works in these areas and position our proposed modular KDN-based framework within the existing literature. In \cite{s20195603}, the authors introduce a four-layer IIoT architecture that leverages fog and edge computing along with DDS-based middleware to enable scalable communication and real-time data processing. Their design outlines clear roles for each layer, unifies the address space for fieldbus devices, and employs System-on-Chip platforms with co-processors to manage time-sensitive tasks. In \cite{fi13100258}, a multi-layer SDN-based architecture is presented to ensure secure IT/OT convergence under DDoS threats. The system integrates physical devices, SDN controllers, and IIoT services, enabling centralized traffic control and dynamic routing. Performance is evaluated using Mininet simulations under various DDoS scenarios, focusing on throughput and RTT degradation. The study emphasizes the need for adaptive security and ML-assisted mitigation in SDN-enabled industrial networks.

A traffic analysis and intrusion detection framework is proposed in \cite{s25082395}, targeting cyber-physical systems in IIoT environments. The approach combines unsupervised clustering for traffic segmentation with supervised machine learning models for classification. Deployed at the network edge, this setup reduces latency and enables prompt responses to detected threats. In \cite{9080395}, the authors develop a digital twin of the Festo Cyber Physical Factory to simulate six sequential assembly stations. The system tracks products using RFID, validates each task in real-time, and can redirect items back to prior stations if needed. This modular simulation environment supports operational visibility and dynamic process control. The study in \cite{article} proposes a fog computing-based framework to improve IT/OT convergence in manufacturing. It supports real-time decision-making by processing data near industrial endpoints, enabling predictive maintenance via telemetry and ML, and enhancing wireless connectivity across factory floors. The goal is to increase equipment uptime, improve analytics, and enable seamless digital integration. Finally, the authors of \cite{10170421} present an edge-based framework for cyber-resilient IT/OT networks. It combines machine learning-assisted SDN with multi-path routing and failover mechanisms. Using NetFlow metadata and keep-alive checks, the system monitors device and link status, classifies anomalies with a random forest model, and triggers recovery via programmable IT/OT interfaces.

The reviewed works provide significant contributions to IT/OT convergence, addressing real-time processing, security, and scalability through SDN, edge computing, digital twins, and fog architectures. However, these approaches often lack comprehensive autonomy, dynamic adaptability, and context-aware decision-making critical for Industry 4.0. Our proposed modular KDN-based framework overcomes these limitations by introducing a closed-loop architecture with a Telemetry Collector, Knowledge Builder, Decision Engine, and Control Enforcer. By leveraging graph-based knowledge representations and utility-driven optimization, our framework ensures real-time, adaptive, and autonomous control, achieving superior decision latency, control effectiveness, and system stability.

The rest of the paper is organized as follows: Section II presents the proposed system components. Section~III explains the proposed approach. Section~IV gives the simulation setup and performance evaluation results. Finally, Section~V and Section~VI concludes the paper and outlines future research directions.

\section{System Components}
\begin{figure}[h]
  \centering
  \includegraphics[width=0.8\linewidth]{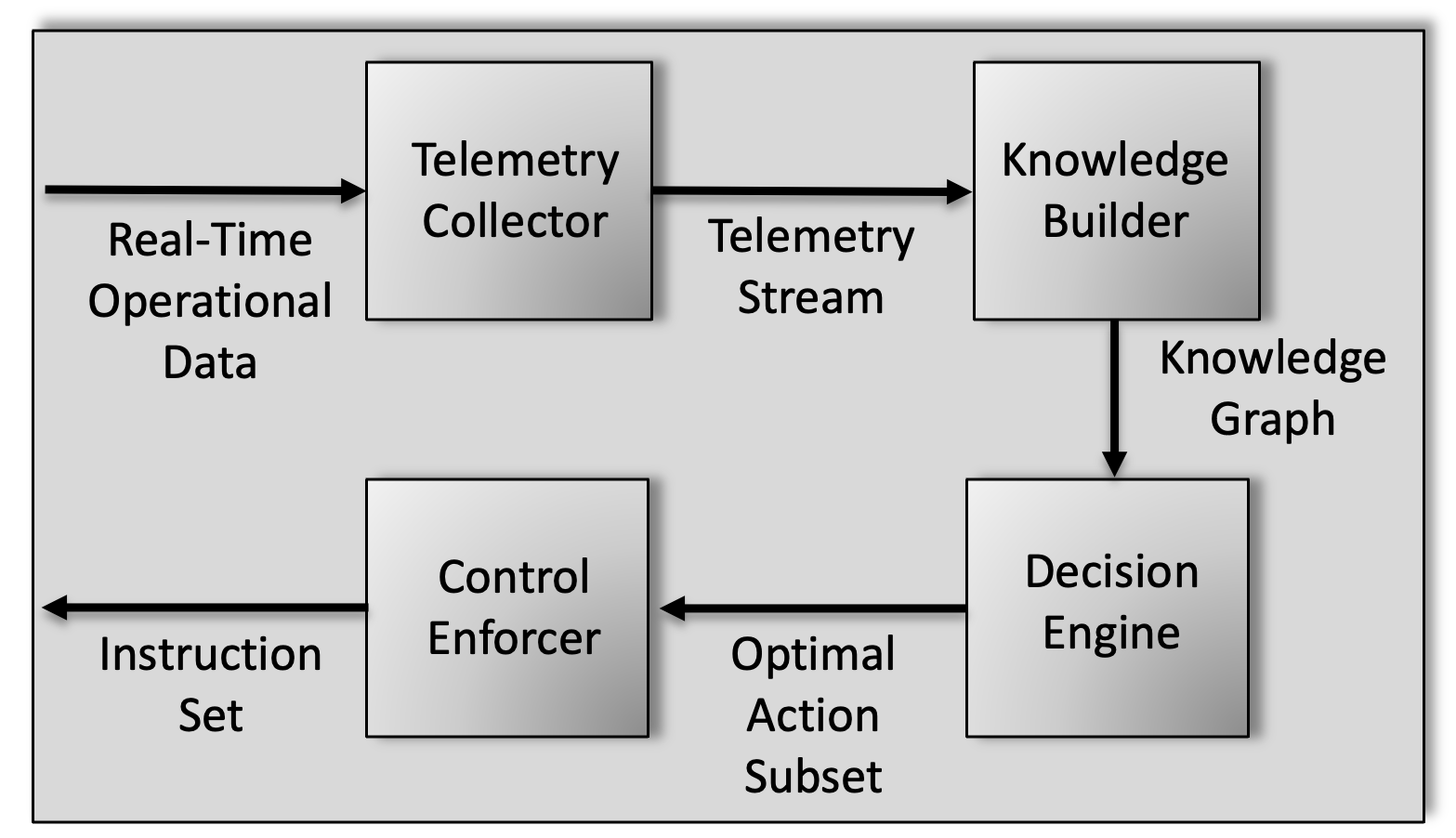}
  \caption{KDN-Based Proposed System}
  \label{fig:architecture}
\end{figure}

To support real-time, autonomous IT/OT integration in complex industrial environments, we propose a modular architecture grounded in the Knowledge-Defined Networking (KDN) paradigm \cite{BILEN2025103984}, \cite{article_1672775}. The architecture introduces a layered intelligence model that enables continuous observation, contextual interpretation, and dynamic control over converged IT and OT domains. Instead of relying on static rule sets or predefined scenarios, the proposed system leverages AI-driven components that learn from evolving conditions and optimize control actions under operational constraints. As illustrated in Fig.~\ref{fig:architecture}, the architecture consists of four main functional modules operating in coordination:
\begin{itemize}
    \item \textit{Telemetry Collector}, which continuously gathers raw data from industrial endpoints, including network traffic patterns, link statistics, queue dynamics, and device-level metrics across IT and OT layers.
    \item \textit{Knowledge Builder}, which transforms this raw telemetry into structured and interpretable knowledge using statistical modeling and graph-based representations. This component captures both system-level behavior and local interactions between nodes.
    \item \textit{Decision Engine}, which applies machine learning and mathematical optimization techniques to determine suitable actions such as path reconfiguration, resource allocation, or load redirection. It accounts for latency, reliability, and capacity constraints critical in industrial use cases.
    \item \textit{Control Enforcer}, which maps high-level decisions to low-level control commands and applies them through SDN controllers and industrial protocol interfaces.
\end{itemize}

Each module operates autonomously yet communicates continuously with the others to maintain a closed-loop control structure. This modular separation enables flexibility, scalability, and ease of integration into heterogeneous industrial networks.

\section{The Proposed Approach}

As described in the system components, the proposed framework introduces a closed-loop control architecture composed of four core modules: \textit{Telemetry Collector}, \textit{Knowledge Builder}, \textit{Decision Engine}, and \textit{Control Enforcer}. These modules interact continuously to enable real-time visibility, semantic abstraction, and autonomous control across converged IT/OT infrastructures. Each module is responsible for a specific transformation layer, forming a pipeline that evolves raw measurements into actionable commands. The details are provided below.

\subsection{Telemetry Collector}
The Telemetry Collector is responsible for the acquisition of real-time operational data from distributed IT and OT entities, such as programmable switches, industrial controllers, and sensor-embedded endpoints. At every time instance $t$, the observation vector for node $i$ is defined as $d_t^{(i)} = (\lambda_t^{(i)}, \delta_t^{(i)}, \gamma_t^{(i)}, \eta_t^{(i)}, \epsilon_t^{(i)})$. Here, $\lambda_t^{(i)}$ represents the link delay, $\delta_t^{(i)}$ the current throughput, $\gamma_t^{(i)}$ the processor load (e.g., CPU usage), $\eta_t^{(i)}$ the queue depth at ingress/egress buffers, and $\epsilon_t^{(i)}$ a binary anomaly indicator triggered by process-level exceptions or threshold violations.

The full telemetry dataset at time $t$ is defined as $\mathcal{D}_t = \{d_t^{(1)}, d_t^{(2)}, \dots, d_t^{(n)}\}$, where $n$ denotes the number of monitored elements. Before being transmitted to the next module, raw data are normalized, time-stamped, and packetized using lightweight edge agents. These agents apply basic preprocessing tasks such as threshold marking, noise filtering, and temporal alignment to ensure coherent aggregation.

\subsection{Knowledge Builder}
The Knowledge Builder converts the telemetry stream $\mathcal{D}_t$ into a semantically enriched and structurally meaningful representation. Specifically, the system constructs a knowledge graph $\mathcal{G}_t = (\mathcal{V}, \mathcal{E}, \mathbf{X})$, where each node $v_i \in \mathcal{V}$ represents a monitored asset such as a machine, edge node, or network switch, and each edge $(v_i, v_j) \in \mathcal{E}$ denotes a direct dependency, interaction, or data exchange between two components. Also, the node feature matrix $\mathbf{X} \in \mathbb{R}^{|\mathcal{V}| \times f}$ encodes recent temporal behaviors of each asset, where $f$ is the number of telemetry attributes (e.g., delay, throughput, load, queue length, alerts). For each node $v_i$, the feature vector $\mathbf{x}_i \in \mathbb{R}^f$ is computed by aggregating the latest $k+1$ measurements across time using a moving average, as shown in Eq.~\ref{e1}. Here, $d_\tau^{(i)}$ denotes the multi-dimensional telemetry vector of node $i$ at time $\tau$. This temporal smoothing emphasizes dominant trends while mitigating transient noise.

\begin{equation} \label{e1}
\mathbf{x}i = \frac{1}{k+1} \sum{\tau = t-k}^{t} d_\tau^{(i)} \quad , \quad k \in \mathbb{Z}^+
\end{equation}

The resulting graph structure $\mathcal{G}_t$ provides a unified representation that captures both the operational behavior (through $\mathbf{X}$) and inter-component relationships (through $\mathcal{E}$). By transforming raw numeric telemetry into a structured, relational format, the system enables high-level reasoning tasks such as dependency inference, state clustering, anomaly localization, and control intent generation. In this context, the term knowledge refers to this graph-level abstraction that embeds contextual, relational, and temporal semantics, which are not directly observable from individual raw measurements alone.

\subsection{Decision Engine}
The Decision Engine takes the system graph $\mathcal{G}_t$ as input and evaluates a finite set of predefined candidate actions $\mathcal{A} = {a_1, a_2, \dots, a_m}$. Each action $a_j$ corresponds to a deterministic infrastructure intervention, including flow redirection, bandwidth reallocation, queue management, or task offloading. To determine the most suitable actions, a utility-driven optimization is performed. For each candidate action $a_j$, two scores are computed as $U(a_j, \mathcal{G}_t)$ and $C(a_j, \mathcal{G}_t)$. Here, $U(a_j, \mathcal{G}_t)$ is the estimated system-level benefit when applying $a_j$ to the current state $\mathcal{G}_t$. This includes measurable gains such as reduced packet delay, improved throughput stability, or decreased resource congestion. $C(a_j, \mathcal{G}_t)$ is the associated operational cost of applying $a_j$, including configuration overhead, state transition latency, and risk of temporary instability.

The optimal subset of actions $\mathcal{A}^*$ is selected by maximizing the net utility under a cost penalty, formulated as given in Eq. \ref{e2}.

\begin{equation} \label{e2}
\mathcal{A}^* = \arg\max_{\mathcal{A}' \subseteq \mathcal{A}} \left[ \sum_{a_j \in \mathcal{A}'} U(a_j, \mathcal{G}_t) - \beta \cdot C(a_j, \mathcal{G}_t) \right]
\end{equation}

In Eq. \ref{e2}, $\beta \in \mathbb{R}^+$ is a scalar that adjusts the trade-off between performance improvement and operational cost. This formulation ensures that selected actions not only maximize benefit but also respect system constraints and dynamic risk profiles. The resulting set $\mathcal{A}^*$ is forwarded to the Control Enforcer for execution.

\subsection{Control Enforcer}
The Control Enforcer module is responsible for converting each selected action $a_j \in \mathcal{A}^*$ into a set of executable low-level commands. This transformation is governed by a deterministic mapping function defined as given in Eq. \ref{e3}.

\begin{equation} \label{e3}
\pi(a_j) = { \texttt{cmd}_1^{(j)}, \texttt{cmd}_2^{(j)}, \dots, \texttt{cmd}_r^{(j)} }
\end{equation}

Each command $\texttt{cmd}_r^{(j)}$ is designed to interact directly with a control interface. For instance, network-related actions are encoded as OpenFlow rules for SDN-enabled switches, while OT-specific instructions are dispatched through industrial communication protocols such as OPC-UA or Modbus, depending on the target device class. Upon execution, the system waits for acknowledgments or error responses from the affected components. Execution latency and feedback status are logged and forwarded upstream to the telemetry layer for post-action analysis.

To support autonomous operation, the architecture employs a real-time feedback loop . The updated telemetry set $\mathcal{D}_{t+1}$ captures the network state after enforcement and is immediately reintegrated into the control cycle. This closed-loop mechanism enables continuous policy refinement, runtime anomaly correction, and system-level performance stabilization without manual reconfiguration \cite{BILEN2020101133}.

\section{Performance Evaluation}

\subsection{Simulation Environment and Experimental Setup}
To evaluate the effectiveness of the proposed modular KDN-based framework, we built a discrete-event simulation environment designed to reflect realistic IT/OT integration scenarios in industrial systems. The simulated topology includes 50 interconnected nodes representing a mix of industrial assets and IT infrastructure components. These nodes consist of programmable logic controllers, process sensors, edge servers, and enterprise-level service nodes operating in coordination. Communication is handled through a hybrid channel model that includes both real-time fieldbus protocols and conventional IP-based networking for data and control traffic.

During each simulation cycle, every node produces telemetry reflecting its operational state, including latency, throughput, queue occupancy, and CPU load. This data is collected by the Telemetry Collector module and processed into a time-aligned system state by the Knowledge Builder, which constructs a graph-based representation at ten-millisecond intervals. The Decision Engine interprets this graph to determine which control actions will maximize system efficiency under current conditions. Actions include rerouting flows, reprioritizing links, and reallocating tasks. The Control Enforcer then translates these high-level decisions into executable control commands using SDN-compatible interfaces and industrial control APIs. Simulations were conducted over 1000 time steps under varying network conditions, including load fluctuations, link degradation, and random component failures. To assess the performance of the system, we focused on three primary evaluation metrics. The details of these metrics could be summarized as follows: 

\begin{itemize}
  \item {Control Effectiveness}: The normalized performance gain achieved after action enforcement, measured in terms of throughput stabilization and delay reduction.
  \item {System Stability Score}: The variance of key operational parameters (e.g., queue sizes, CPU load) over time, indicating how well the system maintains equilibrium under dynamic changes.
\item {Decision Latency-}: The average time elapsed between anomaly detection and corresponding action dispatch.
\end{itemize}

\subsection{Comparative Evaluation Based on Key Metrics}
We compare the performance of our proposed KDN-based framework against the following three baseline approaches:

\begin{itemize}
  \item {Threshold-Based Event Triggering (TET)}: A traditional approach where actions are executed when metric values exceed predefined thresholds.
  \item {Heuristic Rule Scheduler (HRS)}: A rule-based method applying static scheduling decisions based on system heuristics.
  \item {Reinforcement Learning-Based Controller (RLC)}: A model-free RL agent trained to select control actions through environment interaction and reward feedback.
\end{itemize}

\begin{figure}[h]
\centering
  \includegraphics[width=0.45\textwidth]{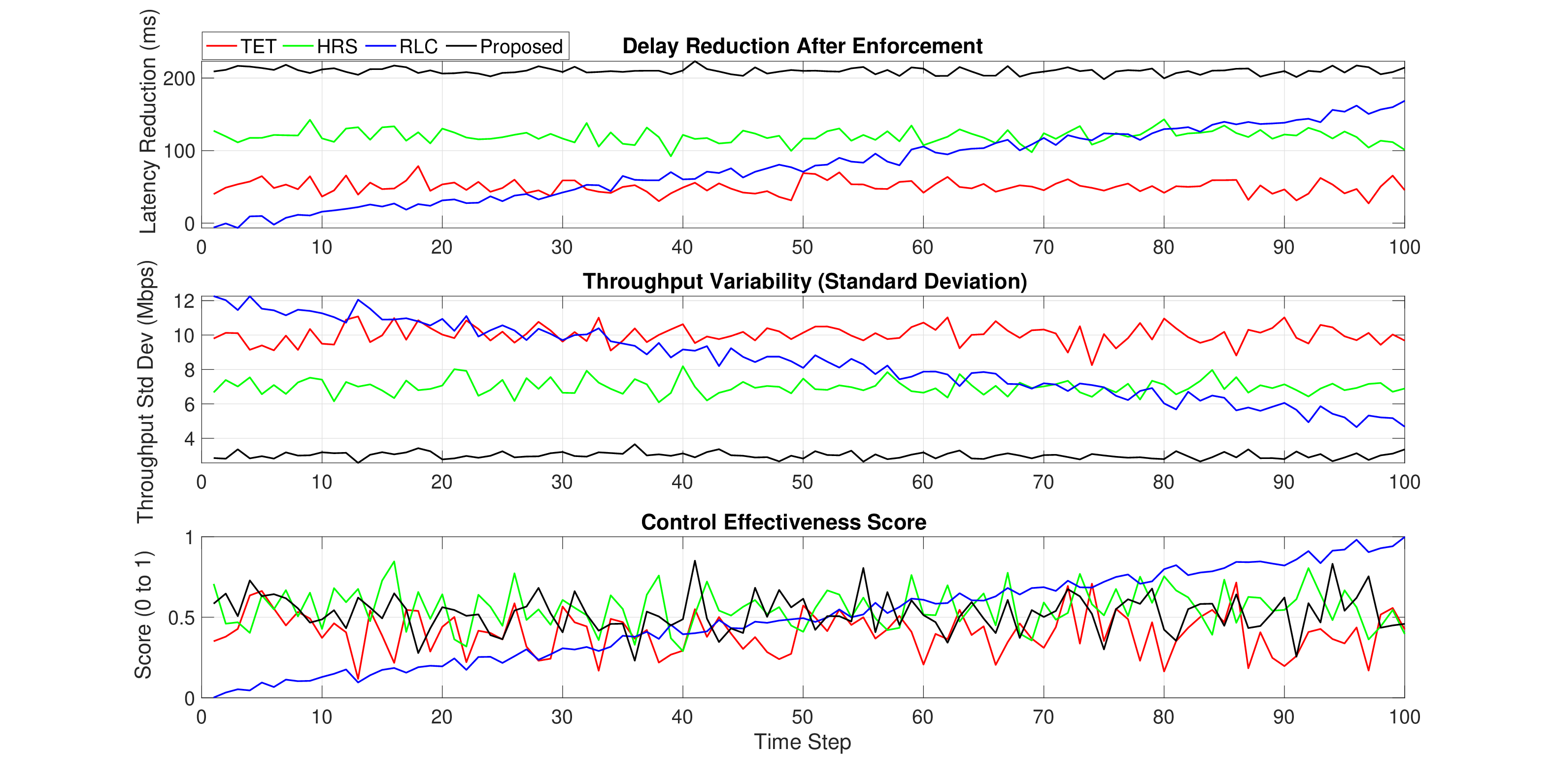}
  \caption{Evaluations for control effectiveness.}
  \label{g1}
\end{figure}

As shown in Fig.~\ref{g1}, the evaluation of control effectiveness is broken down into two sub-metrics: delay reduction after action enforcement and throughput variability, followed by an aggregated Control Effectiveness Score. These subgraphs allow for a detailed analysis of how each method performs in terms of both instantaneous improvement and overall operational stability. In the Delay Reduction subplot, it achieves a sharp and early decline in average latency following action enforcement, demonstrating rapid responsiveness and intelligent action targeting. In the Throughput Variability subplot, the framework maintains stable service delivery with minimal fluctuations, indicating robust stabilization and effective prioritization of actions. The combined Control Effectiveness Score confirms this dual advantage, illustrating consistent superiority across dynamic operational conditions. Here, TET yields the lowest effectiveness. It activates only when predefined thresholds are crossed, often resulting in delayed responses and overcorrections, especially under fluctuating load. These effects are clearly visible in both the delay and variability subgraphs, where TET shows erratic behavior and negligible long-term improvement. HRS shows slightly better control compared to TET, benefiting from domain-specific rules. However, its inflexible rule base leads to static responses that fail to align with evolving system states. As seen in the subplots, HRS achieves moderate delay improvements but struggles with maintaining low variability, causing its overall effectiveness to plateau early. RLC demonstrates higher effectiveness in the later stages of the evaluation. Its learning ability allows it to gradually discover better policies, reflected in improving scores in the final segments of each subplot. However, RLC suffers from cold-start exploration noise and initial policy randomness, which hinder its early responsiveness and stability. In contrast, the proposed method combines instantaneous feedback processing, semantic abstraction, and utility-guided optimization to deliver both early impact and stable performance. This holistic capability is clearly reflected across all three subgraphs in Fig.~\ref{g1}, validating its robustness and adaptability under diverse network conditions.

\begin{figure}[h]
\hspace{-0.2in}  \includegraphics[width=0.45\textwidth]{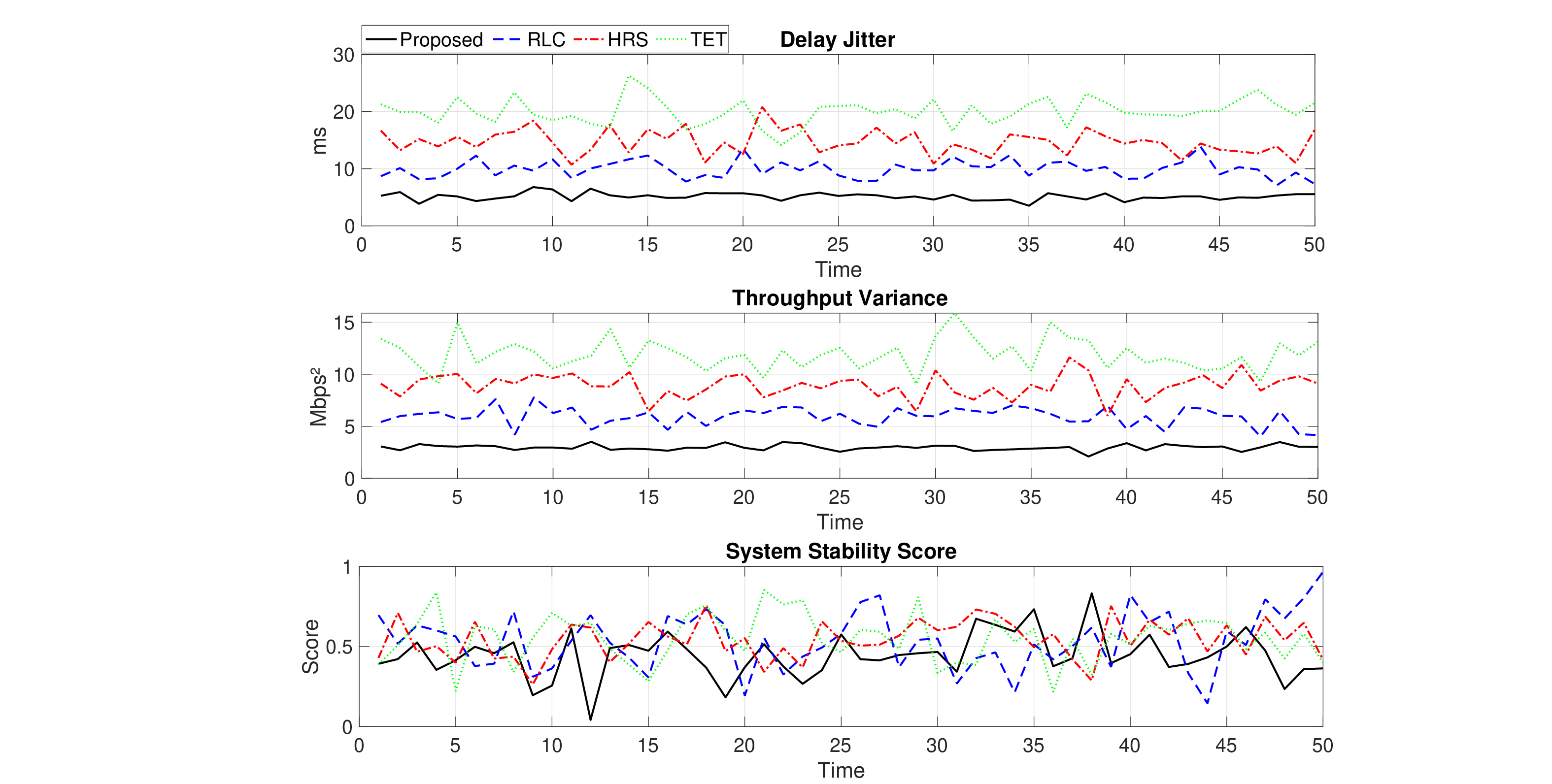}
  \caption{Evaluations for stability score.}
  \label{g2}
\end{figure}

Also, Fig.~\ref{g2} presents a detailed comparison of system stability across methods using three sub-metrics: (i) delay jitter, (ii) throughput variance, and (iii) their composite stability score. Lower jitter and variance indicate smoother and more predictable network behavior, while the overall score combines both metrics to assess temporal consistency. Our proposed method achieves consistently lower jitter and throughput variance, resulting in higher stability scores throughout the evaluation window. This is primarily due to the framework’s adaptive decision-making and context-aware reasoning, which minimizes abrupt shifts in control behavior. In contrast, the RLC method suffers from exploration-induced fluctuations during early episodes. Although its performance improves over time, it remains sensitive to reward shaping and delayed feedback. The HRS method exhibits moderate stability due to its predefined structure but lacks responsiveness to dynamic shifts. Finally, the TET approach performs the worst, frequently overreacting to threshold breaches and causing instability spikes. These results highlight that static or threshold-based methods either lack agility or overreact to transients, while model-free RL requires long convergence times. In contrast, our knowledge-guided strategy ensures immediate and stable adaptation to evolving IT/OT conditions. These results confirm that our modular framework offers faster, more efficient, and stable IT/OT control in complex industrial environments.

\begin{figure}[h]
\centering
  \includegraphics[width=0.45\textwidth]{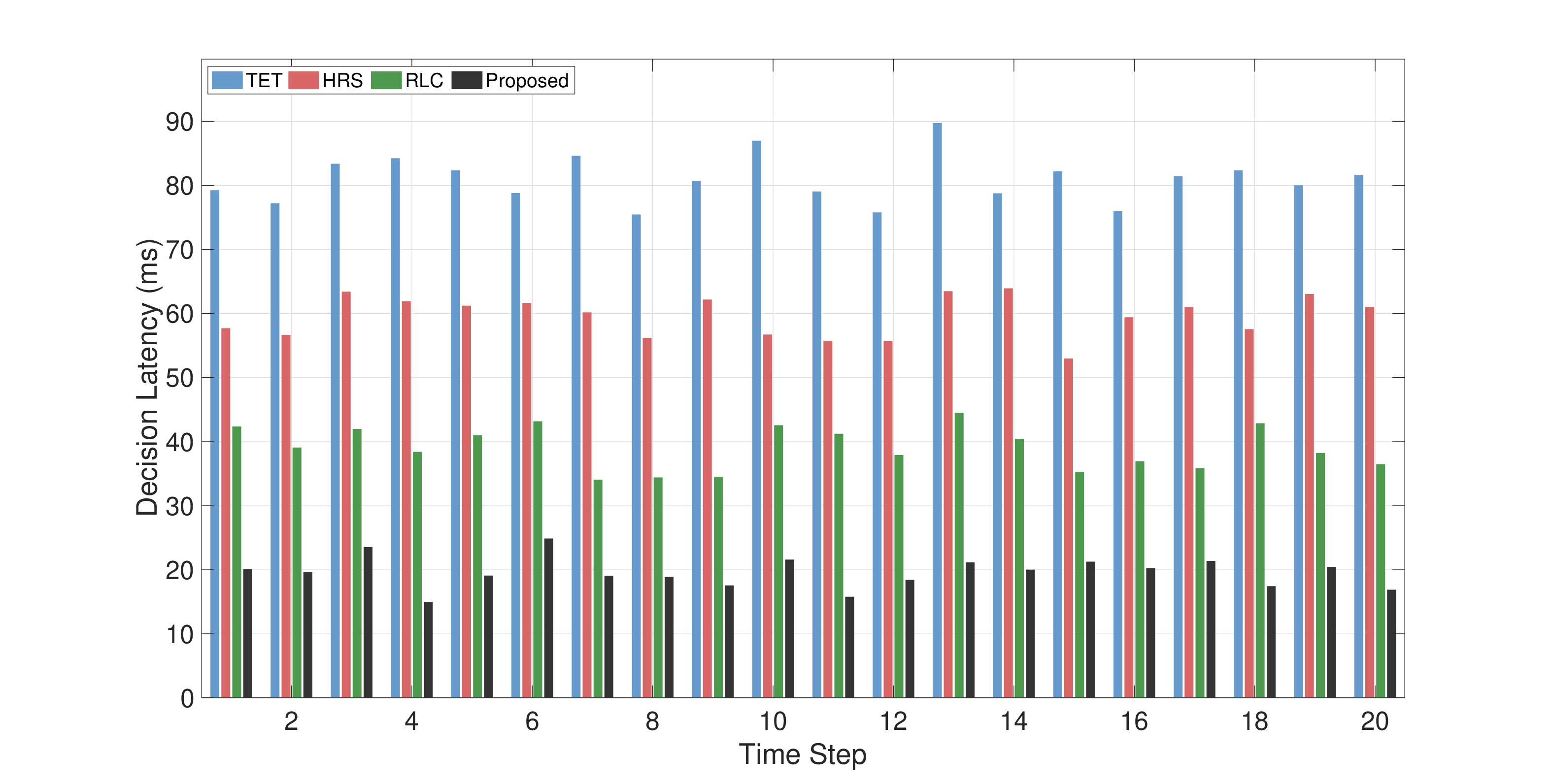}
  \caption{Evaluation for decision latency.}
  \label{g3}
\end{figure}

As explained above, decision latency quantifies the average time taken between the moment a system anomaly is detected and the Decision Engine dispatches the corresponding control action. As the proposed KDN-based architecture employs a closed control loop with modular separation of observation, knowledge extraction, and decision-making, it enables faster and context-aware responses to dynamic conditions. In particular, the use of graph-based system abstraction allows the Decision Engine to reason over dependencies and constraints efficiently, while the utility-optimization mechanism reduces the time spent evaluating alternative actions. As shown in Fig. \ref{g3}, our framework achieves significantly lower decision latency compared to baseline rule-based or threshold-driven systems (e.g., RLC, HRS), which often require manual tuning or lack real-time situational awareness. This improvement demonstrates the advantage of integrating knowledge-driven adaptation into industrial control loops.

\begin{figure}[h]
\centering
  \includegraphics[width=0.45\textwidth]{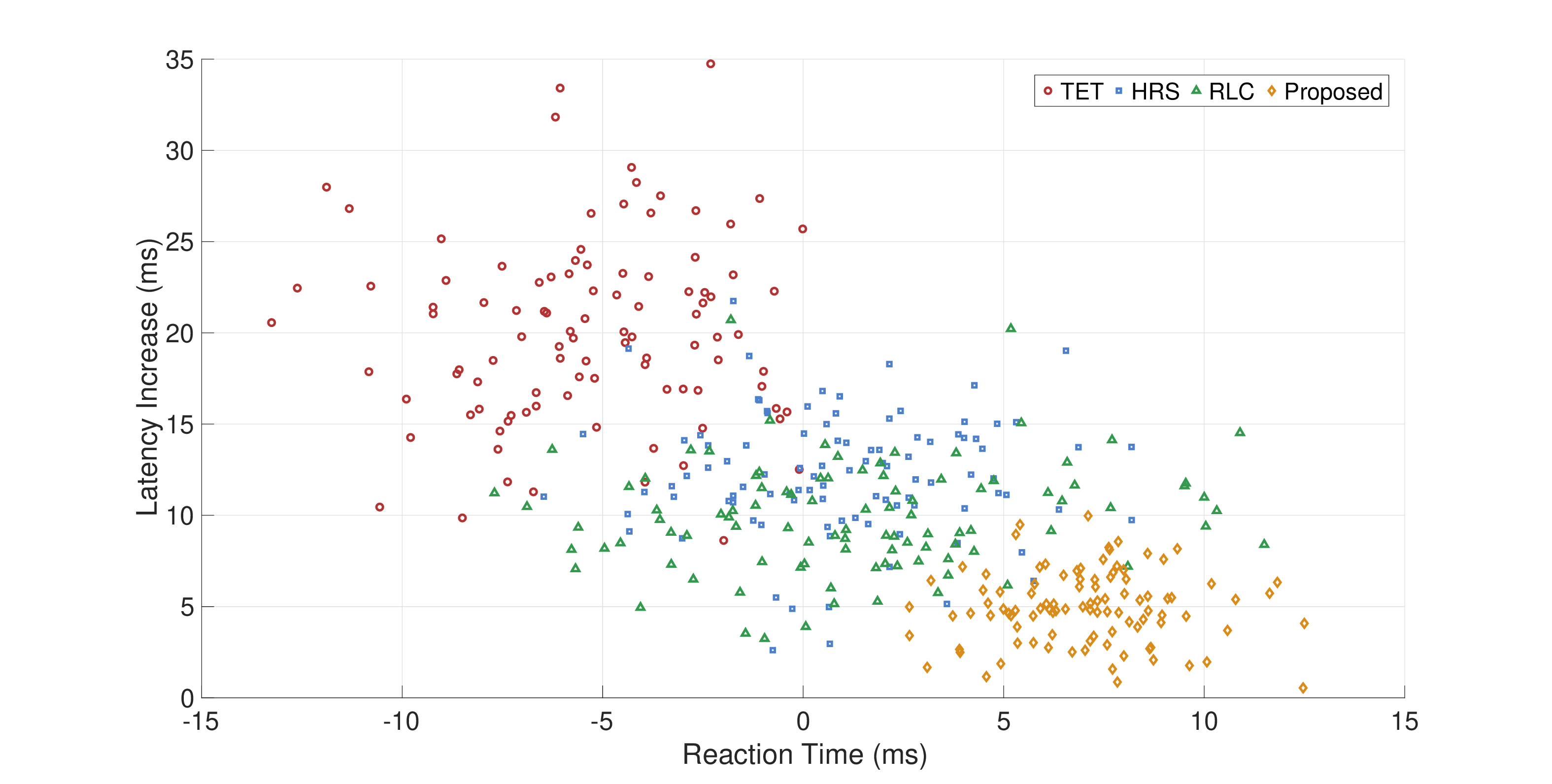}
  \caption{Evaluation for latency increase.}
  \label{g4}
\end{figure}

Also, Fig.~\ref{g4} illustrates that the proposed method reacts faster and incurs less delay increase compared to all baselines. While TET and HRS respond too late or rigidly, RLC shows improvement over time but remains unstable early on.

\section{Conclusion}
In this paper, we proposed a modular, AI-driven framework for enabling autonomous IT/OT integration in modern industrial environments. Our architecture, built upon the Knowledge-Defined Networking (KDN) paradigm, introduced a closed-loop control pipeline that continuously collects telemetry, builds a context-aware system representation, selects optimal actions, and enforces control policies in real-time. Experimental formulation of each module including telemetry modeling, graph-based knowledge abstraction, and utility-based decision optimization demonstrated the feasibility of general-purpose, adaptive control without reliance on extensive retraining or manual configuration.

\section{Future Work} 
While the current framework provides a solid foundation for autonomous management, several extensions are envisioned to enhance long-term adaptability and predictive capabilities. First, we aim to incorporate a lightweight {Digital Twin Layer} that maintains a continuously updated virtual model of the physical infrastructure. This twin would enable short-term forecasting of system states, supporting proactive decision-making before anomalies manifest. Second, we plan to integrate semantic reasoning and abstraction through graph-based learning, enabling the system to detect latent dependencies and learn implicit control patterns from evolving graph structures. Lastly, we will explore {zero-shot policy selection} techniques to allow the system to handle previously unseen scenarios by leveraging semantic similarity, thus reducing the need for explicit retraining in dynamically changing industrial contexts.

\bibliographystyle{IEEEtran}
\bibliography{ref}

\end{document}